\documentclass{article}
\usepackage[utf8]{inputenc}
\usepackage{indentfirst}
\usepackage{epsfig}
\usepackage{amsmath}
\usepackage{authblk}
\usepackage{siunitx}
\usepackage{epsf}
\usepackage{graphicx,xcolor,overpic,mathtools}
\usepackage{amssymb}
\usepackage{multirow,lscape,array}
\usepackage{fullpage}
\usepackage{cite}
\usepackage{float}
\usepackage{enumitem}
\usepackage{amsfonts}
\usepackage{graphicx}
\usepackage{array}
\usepackage{caption}
\newcolumntype{P}[1]{>{\centering\arraybackslash}p{#1}}
\usepackage[colorlinks,linkcolor=blue,anchorcolor=red,citecolor=blue]{hyperref}
\usepackage{cite}

\begin{document}

\title{\bf Analytical proxy to families of numerical solutions: \\ the case study of spherical mini-boson stars}
\author{Jianzhi Yang\footnote{jianzhi@ua.pt}, Pedro V. P. Cunha\footnote{pvcunha@ua.pt}, Carlos A. R. Herdeiro\footnote{herdeiro@ua.pt}
}
\affil{\small \textit{Departamento de Matem\'atica da Universidade de Aveiro and} \\ 
\textit{Centre for Research and
Development in Mathematics and Applications (CIDMA),} \\ \textit{Campus de Santiago, 3810-193
Aveiro, Portugal}}
\date{May 2024}

\maketitle
\begin{abstract}
The Einstein field equations, or generalizations thereof, are difficult to solve analytically. On the other hand, numerical solutions of the same equations have become increasingly common, in particular concerning compact objects. Whereas  analytic approximations to \textit{each individual} solution within a numerical family have been proposed, proxies for \textit{whole} families are missing, which can facilitate studying properties across the parameter space, data compression and a wider usage of such solutions. In this work we tackle this need, proposing a simple strategy based on two different
expansions of the unknown functions in an appropriately chosen basis, to build such proxy. We use as an exploratory case-study spherical, fundamental mini-boson stars, to illustrate the feasibility of such an approach, emphasise its advantage in reducing the data size, and the challenges, say, in covering large parameter spaces.  
\end{abstract}

\section{Introduction}
Non-linear field theories, such as General Relativity (GR), pose a considerable technical challenge in finding both their equilibrium solutions and, even more so, their dynamical ones. As such, closed form solutions are often idealized and a great deal of progress towards more general (and realistic) solutions has been leveraged on computers and numerical methods. In the case of GR, numerical relativity became a central technique to obtain physical results, so important, for instance for the interpretation of gravitational wave transients~\cite{LIGOScientific:2016aoc}. In the realm of equilibrium solutions describing compact objects, remarkably, the black hole solutions of electrovacuum GR are known in (simple) closed forms~\cite{Townsend:1997ku};  on the other hand, neutron stars, black holes beyond electrovacuum, or exotic compact objects are often described as numerical solutions of some Einstein-type field equations - see e.g.~\cite{Schunck:2003kk,Friedman:2013xza,Herdeiro:2015waa,Dias:2015nua}.

Dealing with numerical solutions, rather than analytical ones, adds (at least) two extra hurdles to their generalized usage. The first is to obtain the required data; whereas a closed form solution can generically be found in a usable form in the literature, a numerical solution is often not explicitly given. The second one, is to use the data for the required purposes; whereas any GR practitioner was trained in analysing closed form solutions, examining solutions given as, say, a set of numbers on a grid may not be as straightforward, especially when trying to examine properties along a parameter space.

Naturally, analytic approximations to numerical solutions have been proposed, e.g.~\cite{Cardoso:2014rha,Johannsen:2011dh,Rezzolla:2014mua,Mukazhanov:2024rka,Pappas:2023mbp,Konoplya:2020hyk,Konoplya:2016jvv,Li:2023kqg,Kokkotas:2017ymc,Sajadi:2020axg,Rosa:2022toh}. To the best of our knowledge, however, these have focused on individual solutions within a family, typically spanned by (say) $m$ non-trivial parameters.

To be concrete, imagine the Kerr solution was only known numerically. The available methods provide a ``fit"  for any \textit{specific} value of the spin parameter -- the only non-trivial parameter of the family in this case -- and study the corresponding geometry; but it would be desirable to have \textit{one fit for the whole  1D family} (since the mass is just a scale), enabling a continuous study of the variation of its physical and mathematical properties along the 1D family of solutions. 
 We do note that rare instances exist where multiple solutions can be approximated at once. For example, in \cite{Annulli:2020lyc}, a particular scaling symmetry in the Newtonian branch of Boson Stars (i.e., in the low mass limit) allows covering that branch with a single solution (see also~\cite{Schive:2014dra} for a discussion of a fit in the non-relativistic limit). However, these cases are the exception rather than the rule, making it beneficial to have an approximation framework that can be applied more generally.

In this paper we propose a simple first step towards such a goal. For simplicity we focus on 1D family of solutions, that is, which can be faced as labelled by a single parameter that changes its physical properties, besides any other parameters that may fix scales only. {We show how to perform a double expansion of the unknown functions on a given chosen basis}, to obtain a proxy for the solutions along the 1D family of solutions, or a subset thereof. For concreteness we focus on the family of static and spherical, fundamental, mini-Boson Stars (BSs)~\cite{Kaup:1968zz,Ruffini:1969qy} which over the last decades have become a widely used model for an exotic compact object (see~\cite{Schunck:2003kk,Liebling:2012fv,Jetzer:1991jr} for some recent reviews). We will show that the constructed proxy can provide an approximation with a maximal violation of the field equations of order $\sim 10^{-5}$ for a chosen subset of the BS family. Moreover, there is a considerable gain in the storage space of the proxy data vs. that of the numerical solutions directly obtained from a solver. Still, one observes that it is challenging to obtain a one-fit-all proxy for the \textit{full} parameter space of the family of solutions.

The structure of this paper is outlined as follows: Section \ref{secApprox} details two methodologies employed to attain analytic approximations of differential equations. Section \ref{secBS} introduces the spherical mini-BSs family of solutions, that will be our case-study in this paper. Section \ref{sec:ApproxBS} is dedicated to presenting analytic approximations to a single BS, comparing the two methodologies introduced and considering also how the proxy performs as a solution of the field equations. Then, in Section~\ref{sec:ApproxFamily} this analysis is then expanded to encompass a whole range of solutions within the BS family.  Finally, Section \ref{sec:conclusions} contemplates prospective avenues for future inquiry.

\section{Analytic approximations}\label{secApprox}
We have in mind a spherical solution of the Einstein-matter equations or some generalization thereof. Both the geometry and the matter fields will be specified in terms of a certain number of real functions $f(x)$ and we can conveniently compactify the radial coordinate into an $x$-coordinate that takes values in the unit interval. 

In order to build our proxy, in this section we consider one real scalar function \(f(x)\), defined in the interval \(x \in[0,1]\), which we aim to approximate by some analytical expression. For this purpose, we shall present two different methods: the ``Standard'' and the ``Acceleration-Informed'' methods, detailed below.

\subsection{Standard approximation method}
The original function \(f(x)\) can be approximated by a new function \(\widetilde{f}(x)\), which is a polynomial expansion up to order ${N-1}$:

\begin{equation}
    f(x) \simeq \widetilde{f}(x) =\sum_{i=0}^{{N-1}} \alpha_i\,\mathcal{T}_i(x),
\end{equation}
where $x\in[0,1]$ and $\alpha_i\in\mathbb{R}$. The basis functions $\{\mathcal{T}_i\}$ are adapted Chebyshev polynomials, which are defined as $\mathcal{T}_i(x) \equiv T_i(2x-1)$. The functions $T_i(\chi)$, where $\chi\in[-1,1]$, are the usual Chebyshev polynomials of the first kind~\cite{Chebyshev:1854}.

The polynomials $\mathcal{T}(x)$ satisfy the following recurrence relations, inherited from the standard Chebyshev polynomials~\cite{GeraldWheatley:2003}:
\begin{equation}
\begin{aligned}
    & \mathcal{T}_0(x)=1, \\
    & \mathcal{T}_1(x)=2x-1, \\
    & \mathcal{T}_{k+1}(x)=2 (2x-1) \mathcal{T}_k(x)-\mathcal{T}_{k-1}(x) \text{ for } k \geq 1\,,\quad k\in\mathbb{N}.\\
\end{aligned}
\end{equation}

Similarly to Chebyshev polynomials, the functions $\mathcal{T}_i$ also form an orthogonal basis set with respect to the inner product:
\begin{equation}
   \left\langle\mathcal{T}_i \mid \mathcal{T}_j\right\rangle= \int_{0}^1 \frac{1}{\sqrt{x(1-x)}}\mathcal{T}_i(x) \, \mathcal{T}_j(x) \mathrm{~d} x=
   \left\{\begin{array}{ll}
0 & \text { if } i \neq j \\
\pi & \text { if } i=j=0 \\
\pi / 2 & \text { if } i=j \neq 0
\end{array}\right\}
   =\frac{\pi}{2} \delta_{ij}\left(1+\delta_{i_0}\right).\\
\end{equation}
where $i,j\in\mathbb{N}_0$. In this context, the inner product between two functions $\{u,v\}$ in the domain $x\in[0,1]$ is defined as:
 \begin{equation}\label{eq-inner}
\langle u \mid v\rangle \equiv \int_0^1 \frac{u(x)\,v(x)}{\sqrt{x(1-x)}}\,d x.\\
\end{equation}

Since the original function $f(x)$ and its approximation $\widetilde{f}(x)$ will not typically coincide, it is useful to introduce the residue function $\mathcal{R}(x) \equiv f(x)-\widetilde{f}(x)$. A criteria to minimize the residue $R(x)$ (on average) is to consider the Galerkin method\cite{Galerkin:1915} of weighted residuals:
\begin{equation}
\label{eq1}
\left\langle \mathcal{R} \mid \mathcal{T}_j \right\rangle=0, \qquad \forall\, j \in\{0,1, \cdots, {N-1}\}\,.
\end{equation} 
Equation (\ref{eq1}) then implies:
\begin{equation}
\left\langle f \mid \mathcal{T}_j\right\rangle=\sum_{i=0}^{{N-1}} \alpha_i\langle \mathcal{T}_i\left|\mathcal{T}_j\right\rangle=\alpha_j \,\frac{\pi}{2}\left(1+\delta_{j_0}\right).
\end{equation}
This yields an expression for the basis coefficients $\alpha_j$ of the approximation $\widetilde{f}$:
\begin{equation}
\alpha_j= \frac{2}{\pi\left(1+\delta_{j 0}\right)}\,\left\langle f \mid \mathcal{T}_j\right\rangle,\qquad j \in\{0,1, \cdots, {N-1}\}.
\end{equation}

The standard method detailed above optimizes the approximation of the original function $f(x)$. Despite having a fairly fast convergence, the approximation curve $\widetilde{f}(x)$ will typically oscillate around the original function, with the amplitude of these oscillations tending to decrease as more basis functions are included. This feature is not an issue if the aim is to just approximate the function $f$. However, these oscillations do become a problem if the aim is also to obtain a good approximation for the first and second derivatives of $f$ by directly differentiating $\widetilde{f}(x)$. This issue is particularly relevant if the function $f$ satisfies some second order field equation, as it so often happens in physical systems.\\

Motivated by this problem, in the next subsection we shall introduce a different approximation method, that instead of optimizing the approximation of $f(x)$, it optimizes instead the approximation of the second derivative of $f$, and then reconstructs the original function by integration. This integration procedure enforces a smoothing operation that considerably reduces the oscillation of $\widetilde{f}(x)$. Given that the method primarily uses the information from the second derivative to construct the approximation, it will be termed as an ``Acceleration-Informed'' method, drawing an analogy to Newtonian classical mechanics.\\

\subsection{Acceleration-Informed method for analytic approximation}

Consider that the original function \(f(x)\) satisfies some second-order differential equation in the interval \(x \in[0,1]\), and that we have access to both the values of the function \(f(x)\) and its second derivative \(f^{\prime \prime}(x)\). This will be the case for the metric functions obtained through ordinary differential equation (ODE) numerical solvers of the Einstein equations (or generalizations thereof).

In contrast to the previous methodology, in this approach we shall optimise the approximation of the second derivative $f^{\prime\prime}(x)$, instead of $f(x)$. Then the approximation for $f$ is obtained via integration of the approximated second derivative function. We can as before write the basis expansions for $f$ and its derivatives using adapted Chebyshev polynomials:

\begin{align}
 \label{eq-basis0}
f(x) &\simeq \widetilde{f_0}(x)= \sum^{{N-1}}_{i=0} \alpha_i \,\mathcal{T}_i(x),\\    
\label{eq-basis1}
f^{\prime }(x) &\simeq \widetilde{f_1}(x)= \sum^{{N}}_{i=0} \beta_i \,\mathcal{T}_i(x), \\ 
\label{eq-basis2}
f^{\prime \prime}(x) &\simeq \widetilde{f_2}(x)= \sum^{{N}+1}_{i=0} \gamma_i \,\mathcal{T}_i(x), 
\end{align}
In this method we shall optimize the residue defined via the second derivative, $i.e.$ $\mathcal{R}_{2}(x)\equiv f^{\prime \prime}(x) -\widetilde{f_2} (x)$. Following a procedure similar to the previous method, we can obtain:
\begin{equation}
\gamma_i=\frac{2}{\pi} \frac{1}{\left(1+\delta_{i 0}\right)}\left\langle f^{\prime \prime} \mid \mathcal{T}_i\right\rangle, \qquad i \in \mathbb{N}_0.
\end{equation}
We can relate $\gamma_i$ with $\alpha_i,\beta_i$ via the integration property of the adapted Chebyshev polynomials: 
\begin{equation}
\int \mathcal{T}_k(x) d x=a_k \mathcal{T}_{k+1}(x) +b_k \mathcal{T}_{k-1}(x) + \text { const. },\qquad k \in \mathbb{N}\,.
\end{equation}
In the previous equation, the numbers $a_k$ and $b_k$ are structure constants,
\begin{equation}
    a_k=\frac{1}{2(k+1)(1+[k>0])}\,, \qquad b_k=-\frac{1}{4} \frac{[k \geqslant 2]}{(k-1)}\,,
\end{equation}
where we have used the Iverson logical bracket:
\begin{equation}
\text {  [$P$ }]=\left\{\begin{array}{l}
1 \text { if (Logic relation $P$ is True) } \\
0 \text { if (Logic relation $P$ is False) }
\end{array}\right. .
\end{equation}

From the integration relations, it is possible to obtain the following recurrence relations:
\begin{itemize}
    \item $\quad \beta_i=\gamma_{i-1} a_{i-1}+\gamma_{i+1} b_{i+1}$, \quad with $i\in\{1,\cdots,{N}\}$.
    \item $\quad \alpha_i=\beta_{i-1} a_{i-1}+\beta_{i+1} b_{i+1}$, \quad with $i\in\{2,\cdots,{N-1}\}$.
\end{itemize}
By setting a cut-off $\beta_i=0$ for $i>{N}$, and similarly $\alpha_i=0$ for $i>{N-1}$, the recurrence relations above determine all non-trivial coefficients of $\alpha_i$, except $\alpha_0$ and $\alpha_1$, which are integration constants.
Applying the Galerkin method to minimize the residue of $\mathcal{R}_{0}(x)\equiv f(x) -\widetilde{f_0} (x)$, we can obtain:
\begin{equation}
    \alpha_0=\frac{1}{\pi} \left\langle f \mid \mathcal{T}_0\right\rangle\,,\qquad     \alpha_1=\frac{2}{\pi} \left\langle f \mid \mathcal{T}_1\right\rangle .
\end{equation}

We remark that generically $\widetilde{f_0}^{\prime\prime} \neq \widetilde{f_2}$, due to the tail end cutoff of the basis expansion. However, typically these differences will be small for sufficiently large ${N}$.

\section{Boson Star solutions}\label{secBS}

In 1955, John Wheeler explored whether electromagnetic and gravitational waves could form a particle-like solution in  the context of Einstein's theory of gravity, leading to the discovery of unstable solutions known as \textit{geons}\cite{Wheeler:1955zz, Power:1957zz}. Analogous (in spirit) stable solutions were found in 1968 by D. J. Kaup, which eventually became known as boson stars (BSs), who expanded on Einstein's gravity by integrating a complex scalar field, resulting in what is known as the Einstein-Klein-Gordon theory\cite{Kaup:1968zz} - see also R. Ruffini and S. Bonazzola~\cite{Ruffini:1969qy}. Since then, many models of BSs have been put forward in the literature, invariable obtained as numerical solutions of the corresponding field equations~\cite{Liebling:2012fv, Liang:2022mjo}.

For our purposes here, we shall consider the Einstein-Kein-Gordon theory, with gravity minimally coupled to a complex scalar field  $\Phi$ with mass $\mu$, and with no self-interactions:
\begin{equation}
\label{action}
 S=\int  d^4x\sqrt{-g}\left[ \frac{R }{16\pi G }  
  - \Phi_{, \, a}^* \Phi^{, \, a}  - \mu^2 \Phi^*\Phi 
 \right].
\end{equation} 

The resulting field equations are:
\begin{eqnarray}
\label{E-eq}
  &&
 R_{ab}-\frac{1}{2}g_{ab}R=8 \pi G~T_{ab} , ~~
	\\
	&&
	\label{KG-eq}
\nabla^a\nabla_a \Phi =\mu^2 \Phi ,  
\end{eqnarray}
where 
\begin{equation}
T_{ab} =  
  \Phi_{ ,  a}^{ *}\Phi_{,b } 
 +\Phi_{,b}^*\Phi_{,a} 
- g_{ab} 
[  
 \Phi_{,c}^{*}\Phi^{,c} +\mu^2\Phi^{*}\Phi 
]
\end{equation}
is the
 energy-momentum tensor  of the scalar field.

BS solutions of this model are often termed \textit{mini-BSs}. To obtain the simplest ones, a spherically symmetric, static line element can be described in isotropic coordinates $\{t,r,\theta,\varphi\}$:
\begin{eqnarray}
\label{metric}
ds^2=  -e^{2F_0(r )} dt^2+e^{2F_1(r )} 
\left(
       dr^2+r^2 d\Omega^2
\right) ,
\end{eqnarray}
where $d\Omega^2$ provides standard line element of the 2-sphere. Due to assumed symmetries, the metric functions $\{F_0,F_1\}$ only depend on the radial coordinate $r$, with the range $0\leq r<\infty$. A suitable ansatz for the complex scalar field, with an harmonic frequency $\omega$ is provided by:
\begin{equation}
    \Phi = \frac{1}{\sqrt{4\pi G}}\, \phi(r) e^{-i\omega t}.\label{eq:phiansatz}
\end{equation}

These functions should satisfy some suitable boundary conditions, compatible with regularity at the origin and asymptotic flatness:
\begin{align}
&\left.\partial_r F_0\right|_{r=0}=0 &\left.F_0\right|_{\infty}=0\,  ,\\
&\left.\partial_r F_1\right|_{r=0}=0 &\left.F_1\right|_{\infty}=0\,  ,\\
&\left.\partial_r \phi\right|_{r=0}=0 &\left.\phi\right|_{\infty}=0\,  .
\end{align}

The domain of existence of such spherically symmetric BS solutions is represented in Fig.~\ref{img:BSspace}. BS solutions exist along a spiral curve when represented in a diagram of the total mass $M$ vs its harmonic frequency $\omega$. We shall focus on solutions within the regions $A-B$ highlighted in Fig.~\ref{img:BSspace}, around the configuration with the maximum possible mass, where there is a transition between stability and instability - see $e.g$~\cite{Santos:2024vdm}. In both regions, each BS solution can be uniquely parameterized by the harmonic frequency $\omega$.

\begin{center}
\includegraphics[width=0.6\textwidth]{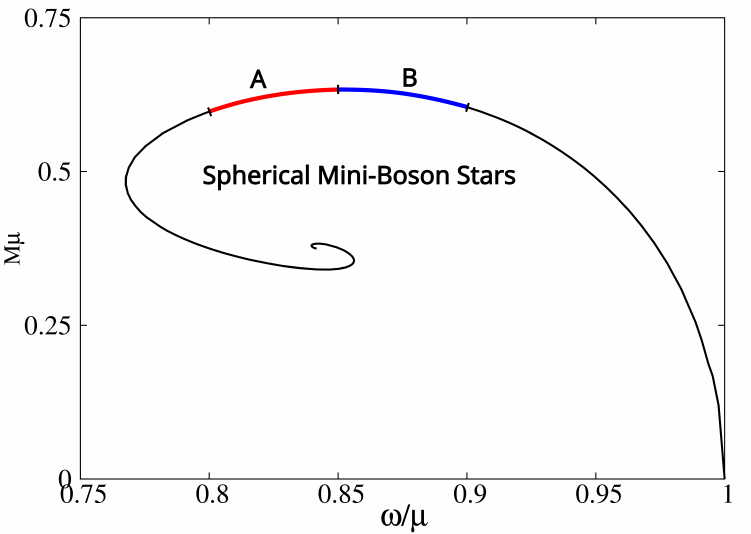}
\captionof{figure}{Domain of existence of static and spherically symmetric BSs, in a diagram of the total mass $M$ of the solution vs its harmonic frequency $\omega$. The different highlighted regions $A$, $B$ will be considered in Section~\ref{sec:ApproxFamily} and are defined in equations~\eqref{Region A} and~\eqref{Region B}.}\label{img:BSspace}
\end{center}
 
In this setting, obtaining a BS solution reduces to solving a set of three ordinary differential equations (ODEs) for the metric functions and the scalar field $\{F_0,F_1,\phi\}$ with the boundary conditions above. These field equations can be explicitly found in \eqref{eq:Ett}-\eqref{eq:EKG}. These ODEs can be solved numerically using standard methods known in the literature, $e.g.$ a shooting method. The raw data of a single BS solution is then usually presented by a discrete set of data points $(r_i, F_0^i, F_1^i, \phi^i)$ along the radial direction, with $i\in\mathbb{N}$. We shall work with units\footnote{The constants $\mu$ and $G$ can be absorbed in the dimensionless code variables and therefore there is no loss of generality in choosing $\mu=4\pi G=1$.} $\mu=1$, $i.e.$ the length scale is set by the Compton wavelength of the bosonic particle. In the following section we shall approximate the data of the BS solution with maximum mass with analytical expressions.

\section{Analytic approximation of a specific Boson Star solution}\label{sec:ApproxBS}

Considering the methodologies presented in Section~\ref{secApprox}, as a first step we aim to formulate an approximation to \textit{a specific} BS solution.
In particular, we shall focus in this section on the BS solution with maximal mass ($M\mu \simeq 0.633$), characterized by the harmonic frequency $\omega/\mu\simeq 0.852626$. 

It will be helpful to introduce a compactified radial coordinate denoted as $x$, defined in the interval $x \in[0,1]$:
\begin{equation}
\label{radial transformation}
x=\frac{r}{1+r}\,.
\end{equation}

This compactification maps the entire radial domain $r\in[0,\infty[$ into the interval $x \in[0,1[$, and so the origin (infinity) is mapped to $x=0$ ($x=1$).
When expressed in this new radial coordinate, the discrete set of solution data points, $\{x_i, F_0^i, F_1^i, \phi^i\}$ with $i\in\mathbb{N}$, can be extended to the continuum through a standard cubic spline interpolation applied to each one of the metric functions individually. For the sake of simplicity, the spline functions will be denoted simply by the associated function label $\{F_0(x), F_1(x), \phi(x)\}$, and are defined in the domain $x\in[0,1]$. For the remainder of this article, we may also adopt the compact notation $F_k$, with $k\in\{0,1,2\}$, as a short-hand for the metric functions and the scalar field $\{F_0, F_1, F_2\equiv \phi\}$.

As detailed in the Section~\ref{secApprox}, $F_k(x)$ can be approximated by some function $\widetilde{F}_k(x)$, expressed as a polynomial basis expansion:
\begin{equation}
  \label{eq-Basis}
F_k(x) \simeq \widetilde{F}_k(x)= \sum^{{N-1}}_{i=0} \alpha^{(k)}_i \,\mathcal{T}_i(x)\, , \qquad k\in\{0,1,2\}\, . 
\end{equation}

The coefficients $\alpha_i^{(k)} \in \mathbb{R}$ above can be found in Appendix~\ref{sec:appendix} for both the ``Standard Method'' and ``Acceleration-Informed Method'' detailed in Section~\ref{secApprox}. For the functions $F_0$ and $F_1$ we will consider polynomials up to ${N}=34$, while for the function $\phi$ the cut-off will be at ${N}=40$. \\

The number of coefficients in both methods might initially appear large, especially in comparison with other approximation methods in the literature, such as the continuous fraction approximation~\cite{Mukazhanov:2024rka, Kokkotas:2017ymc}. The latter is often noted for its rapid convergence with very few coefficients. Existing precision tests for such approximations generally focus on deviations in observational quantities (e.g., the orbital frequency at the Innermost Stable Circular Orbit or the black hole shadow radius) and less frequently on the errors in the actual metric functions or the degree of violation of the underlying field equations. While studies reporting on such violations~\cite{Kokkotas:2017ymc,Li:2023kqg,Mukazhanov:2024rka} use a limited number of parameters, they tend to exhibit higher errors compared to those presented in this paper. The use of continued fractions may potentially lead to faster convergence and could be considered in future extensions of this work. Nevertheless, when performing derivatives and integrations, adapted Chebyshev polynomials may have the advantage of leading perhaps to more straightforward analytical computations compared to continued fractions, even if this might come at the cost of having to work with some additional coefficients.\\

In Fig.~\ref{img:Function comparison} we display the approximation functions $\widetilde{F}_k(x)$ for both the ``Standard Method'' and the ``Acceleration-Informed Method'', as well as the original functions $F_k(x)$, $k\in\{0,1,2\}$, for reference. The difference between the curves in each plot of Fig.~\ref{img:Function comparison} is virtually imperceptible to the naked eye.

\begin{center}
\includegraphics[width=0.5\textwidth]{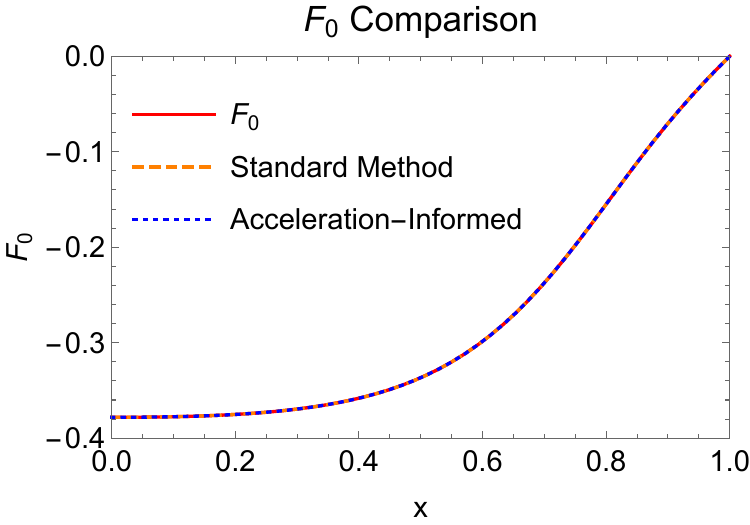}\includegraphics[width=0.5\textwidth]{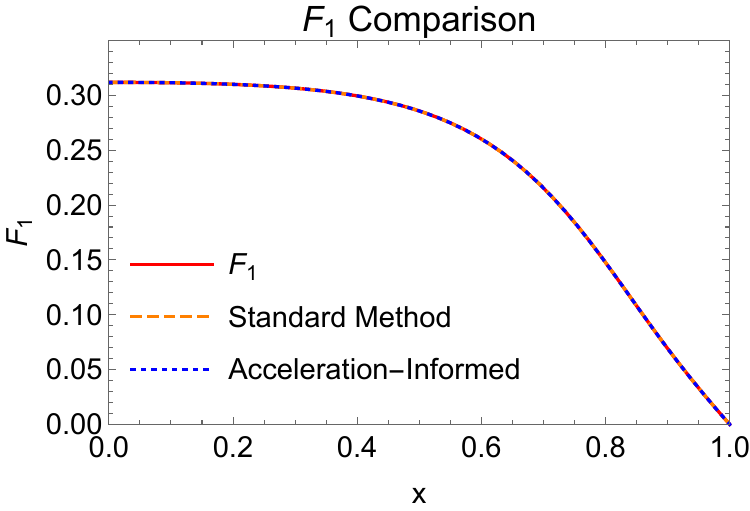}
\includegraphics[width=0.5\textwidth]{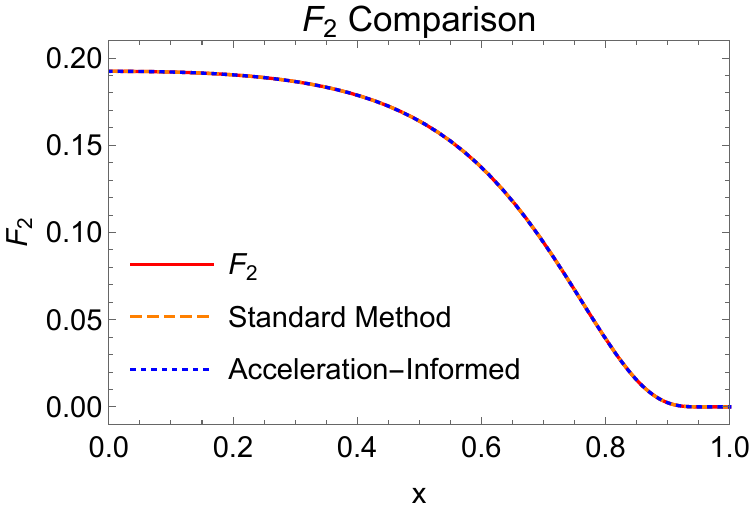}
\captionof{figure}{ Original metric functions $F_k(x)$ and their approximations $\widetilde{F}_k(x)$ with $k\in\{0,1,2\}$, for a BS solution with maximal mass and harmonic frequency $\omega/\mu\simeq 0.852626$. }\label{img:Function comparison}
\end{center}

In order to better compare the error of both approximation methods, we display in Fig.~\ref{img:Error Comparision} the difference between the original function and its approximation (in absolute value), $i.e.$ $\mathcal{E}_k=\left|F_k(x)-\widetilde{F}_k(x)\right|$ with $k\in\{0,1,2\}$, for each one of the methods. From this comparison we can conclude that in terms of accurately approximating the spacetime metric functions the ``Acceleration-Informed method'' outperforms the ``Standard Method''.

\begin{center}
\includegraphics[width=0.5\textwidth]{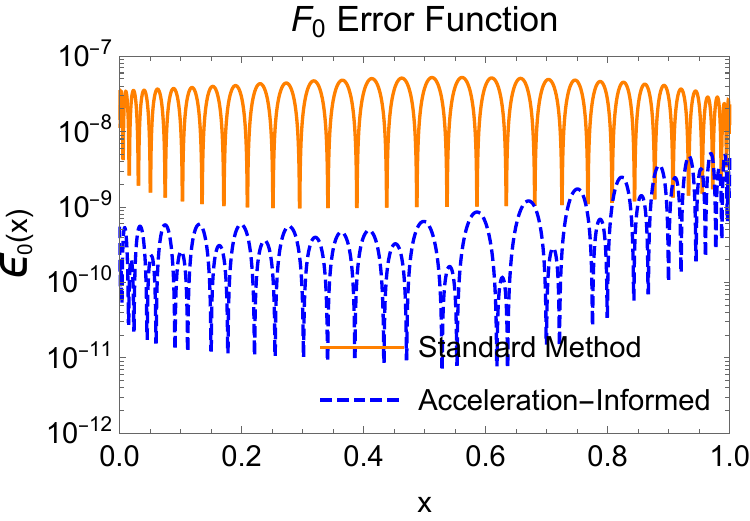}\includegraphics[width=0.5\textwidth]{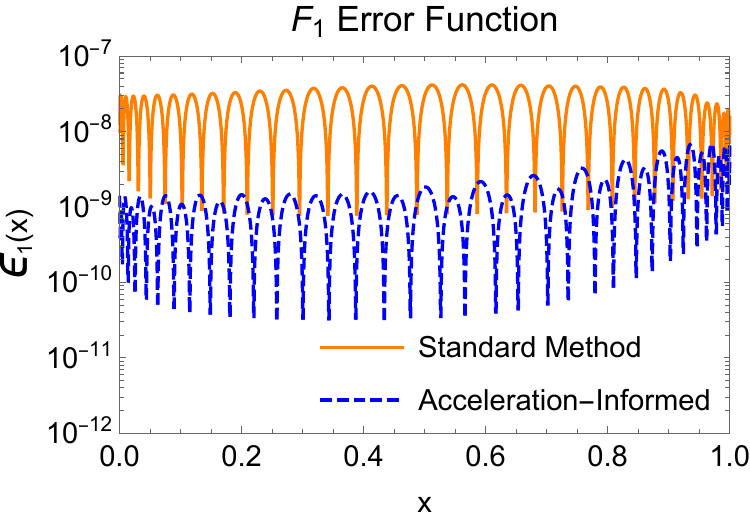}
\includegraphics[width=0.5\textwidth]{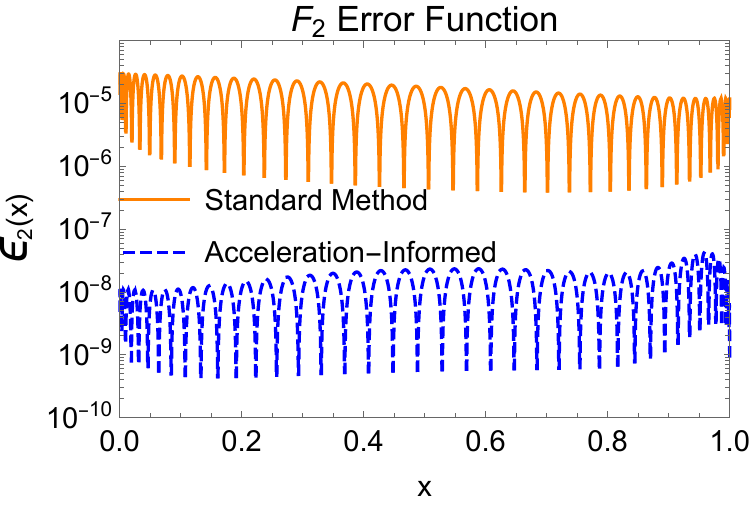}
\captionof{figure}{ Comparison between the approximation errors of the ``Standard Method'' vs the ``Acceleration-Informed Method'', represented in a y-axis logarithmic scale (basis-10), for a BS solution with maximal mass and harmonic frequency $\omega/\mu\simeq 0.852626$.}\label{img:Error Comparision}
\end{center}

How well does the metric approximations solve the initial field equations~\eqref{E-eq} and \eqref{KG-eq}? The non-trivial field equations, when written in terms of the metric functions $F_k(x)$, are obtained by equating the terms $\{\textrm{E}_{tt},\textrm{E}_{rr},\textrm{E}_{\theta\theta},\textrm{E}_{KG}\}$ below to zero:

\begin{align}
\label{eq:Ett}
 \textrm{E}_{tt}(x)&\equiv x\Big(R_{t t}-\frac{1}{2} g_{t t} R - 8\pi G~T_{t t}\Big)  \\
&= -2x\phi^2\,\left(e^{2F_0}\mu^2+\omega^2\right) - (1-x)^4\,e^{2F_0-2F_1}\left(4F_1' + x\,{F_1'}^2 + 2x\,{\phi'}^2 +2x\,F_1''\right) , \nonumber\\
&\nonumber \\
\label{eq:Err}
 \textrm{E}_{rr}(x)&\equiv x\,\Big(R_{r r}-\frac{1}{2} g_{r r} R - 8\pi G~T_{r r}\Big)\\
 &= 2x\phi^2\, e^{2F_1-2F_0}\left(e^{2F_0}\mu^2-\omega^2\right) - (1-x)^3\left\{-2F_0' -F_1'\Bigg(2 + x\,(1-x)(F_1'+2F_0')\Bigg) + 2x\,(1-x){\phi'}^2 \right\},\nonumber\\
 &\nonumber \\
\label{eq:Ethth}
 \textrm{E}_{\theta\theta}(x)&\equiv R_{\theta\theta}-\frac{1}{2} g_{\theta\theta} R - 8\pi G~T_{\theta\theta}\\
&=\frac{2x^2\phi^2}{(1-x)^2}e^{2F_1-2F_0}\,\left(e^{2F_0}\mu^2-\omega^2\right) + x(1-x)(1-2x)\left(F_1'+F_0'\right) + x^2(1-x)^2\left({F_0'}^2 + 2{\phi'}^2 +F_1'' +F_0''\right),\nonumber\\
& \nonumber\\
\label{eq:EKG}
 \textrm{E}_{\textrm{KG}}(x)&\equiv \Big(\sqrt{4\pi G}\,x\,e^{i\omega t}\Big)\Big(\nabla^a\nabla_a \Phi -\mu^2 \Phi\big) \\
 & = -x\,\phi\,e^{-2F_0}\left(e^{2F_0}\mu^2-\omega^2\right) + (1-x)^4\,e^{-2F_1}\Big( \left(2+xF_1' + xF_0'\right)\,\phi' + x\phi''\Big) . \nonumber\\
 & \nonumber
\end{align}

 The expressions above desingularise the field equations at $x=0$, and also remove inconsequential global factors. When the field equations are satisfied, it implies that the terms $\{\textrm{E}_{tt},\textrm{E}_{rr},\textrm{E}_{\theta\theta},\textrm{E}_{KG}\}$ are all equal to zero. By displaying the values of these expressions as functions of the radial coordinate $x$, we can numerically assess the degree to which the metric approximations accurately satisfy the field equations.
 To illustrate these violations, in Fig.~\ref{Fig:KG_methods} we display the numerical values of $\textrm{E}_{\textrm{KG}}(x)$. This function is computed for both the ``Standard'' and ``Acceleration-Informed'' approximations methods, along the original data from the ODE solver. From Fig.~\ref{Fig:KG_methods}, it is clear that the ``Acceleration-Informed method'' significantly outperforms the ``Standard method'' in satisfying the Klein-Gordon Equation, with a typical error of $\sim 10^{-5}$ for this particular BS solution.
 
\begin{center}
\includegraphics[width=0.5\textwidth]{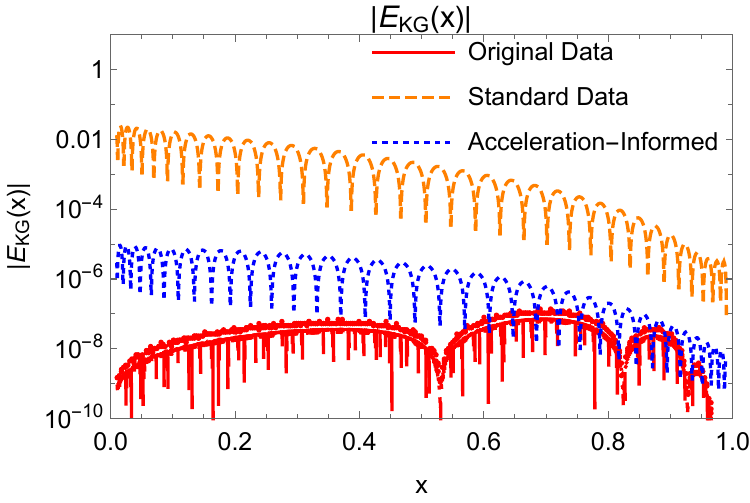}
\captionof{figure}{ Logarithmic plot of the error function $\textrm{E}_{\textrm{KG}}(x)$ (in modulus). This plot includes data obtained from both approximation methods, as well as the original data generated by the ODE solver. }\label{Fig:KG_methods}
\end{center}

Considering only the ''Acceleration-Informed Method'', in Fig.~\ref{Eq(x) comparison} we represent the values of all the functions $\{\textrm{E}_{tt},\textrm{E}_{rr},\textrm{E}_{\theta\theta},\textrm{E}_{KG}\}$, which showcases that the field equations are in general satisfied up to order $\sim 10^{-5}$ for the Klein-Gordon equation, and $\sim 10^{-6}$ for the other Einstein field equations.

\begin{center}
\includegraphics[width=0.5\textwidth]{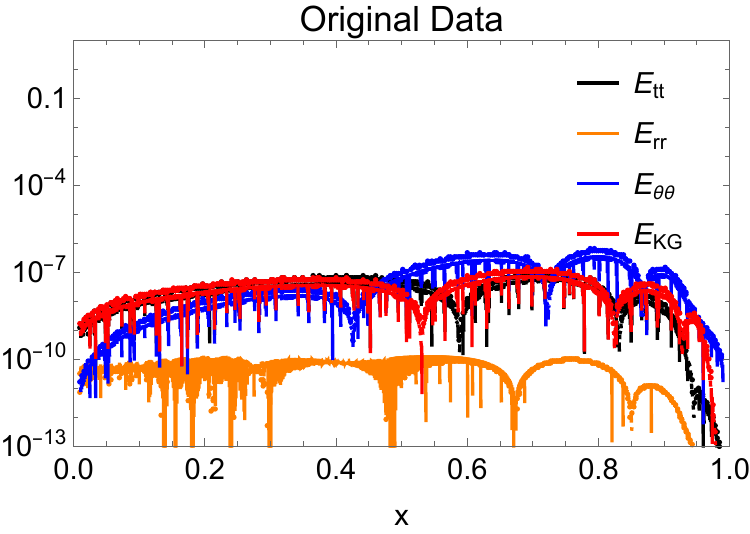 }\includegraphics[width=0.5\textwidth]{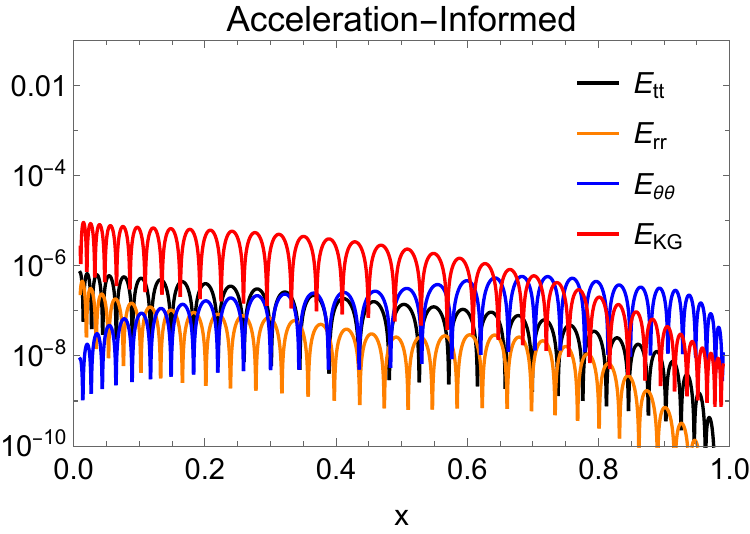 }
\captionof{figure}{Representation of the error functions $\{\textrm{E}_{tt},\textrm{E}_{rr},\textrm{E}_{\theta\theta},\textrm{E}_{KG}\}$ in a logarithmic scale, using the original data (left plot) and the Acceleration-Informed Approximation (right plot).}\label{Eq(x) comparison}
\end{center}

Within the violations reported, the ``Acceleration-Informed'' approximation method attains some level of success in satisfying the field equations for this gravity theory. In this approach, increasing the number of polynomials basis usually leads to more accurate approximations. Nonetheless, due to the existence of truncation round-off errors there is often a limit to how much we can improve the accuracy by rising ${N}$. We identified ${N}=34$ for the functions $\{F_0, F_1$\} and ${N}=40$ for the function $\phi$ as the options that lead to some of the most accurate approximations possible.

\section{Analytic Approximation for a Family of Boson Star Solutions}\label{sec:ApproxFamily}
The previous section focused on the approximation of a single BS solution. How to approximate a {\it family of BS solutions}?

Different BS solutions can be approximated by a different set of coefficients $\alpha_i^{(k)}$, for instance, obtained through the ``Acceleration-Informed approximation method''. As this method is applicable to individual BS solutions, a series of BS solutions, each characterized by a different frequency $\omega$, will yield a function $\alpha_i^{(k)}(\omega)$ within the BS solution spectrum. This function, responsible for providing the approximation coefficients corresponding to each $\omega$, can itself be approximated through a polynomial basis expansion, dependent on the variable $\omega$:

\begin{equation}
\label{eq-phi-basis}
\alpha_i^{(k)}(\omega) \simeq \widetilde{\alpha}_i^{(k)}(\omega)= \sum_{j=0}^{{n-1}} C_{ij}^k \,\psi_j(\omega)\,,
\end{equation}
with $i\in\{0,\cdots, {N-1}\}$, $j\in\{0,\cdots,{n-1}\}$, and $k\in\{0,1,2\}$. In this expansion we have introduced the coefficients $C_{ij}^k\in \mathbb{R}$ with respect to a new polynomial basis $\{\psi_j\}$. By knowing these coefficients, one can approximate all the metric functions for a sequence of BS solutions. In this context, it will be helpful to introduce a mapping $X(\omega)$, to be specified soon, that projects the variable $\omega$ into the auxiliary variable $X\in [0, 1]$.
The explicit form of the basis $\{\psi_j\}$, constructed from adapted Chebyshev polynomials, can then be defined as:
\begin{equation}
    \psi_j(\omega) \equiv \mathcal{T}_j(X(\omega))\,.
\end{equation}

By applying the ``Standard method'' detailed in Section~\ref{secApprox} we can obtain the values of the coefficients $C_{ij}^k$:

\begin{equation}
C_{ij}^k= \frac{2}{\pi\left(1+\delta_{j 0}\right)}\,\left\langle \alpha_{i}^{(k)} \circ X^{-1} \mid \mathcal{T}_j \right\rangle,\qquad j \in\{0,1, \cdots, {n-1}\}.
\end{equation}

For the latter expression, the inner product requires a composite function with the inverse $X^{-1}$ of the map $X$, in order to be consistent with the integration domain of~\eqref{eq-inner}.\\
After knowing these coefficients, the complete approximation for the functions $F_k$ reads:
\begin{equation}
  \label{eq-basis3}
F_k(x,\omega) \simeq \widetilde{F}_k(x)= \sum^{{N-1}}_{i=0}\sum^{{n-1}}_{j=0} C^{k}_{ij} \,\mathcal{T}_i(x)\,\mathcal{T}_j(X(\omega)),  \qquad k\in\{0,1,2\}. 
\end{equation}

As our case study of a (sub-)family of BSs, we shall consider BS data in the two different regions $\{A,B\}$ near the maximum BS mass, and represented in Fig.~\ref{img:BSspace}:
\begin{align}
    \textrm{Region }A:\qquad \omega \in [0.80, 0.85]\,,\qquad X(\omega)&=20\omega -16\,,\qquad  n=7. \label{Region A}\\
    \textrm{Region }B:\qquad \omega \in [0.85, 0.90]\,,\qquad  X(\omega)&=20\omega -17\,,\qquad n=4. \label{Region B}
\end{align}
The value of $n$ chosen for the intervals $A$ and $B$ are the minimum value of $n$ (in each interval) that minimizes in practice the approximation error. Larger values of $n$ will not increase the accuracy of the approximation in a significant way since the error saturates.
The errors in satisfying the field equations in each interval $\{A,B\}$ are displayed through a video analysis, accessible via the provided links~\cite{weblinkA}~\cite{weblinkB}. The video clip displays the error functions $\{\textrm{E}_{tt},\textrm{E}_{rr},\textrm{E}_{\theta\theta},\textrm{E}_{KG}\}$ as a function of $x$, and as the values of frequency $\omega$ changes. In the regions considered, and regardless of variations in the value of $\omega$, the field equations have a maximum error $\sim 10^{-5}$, with typical errors often being much smaller, namely of the order of $\sim 10^{-6}$ or lower.\\

It is relevant to consider what is the {\it total} number $\mathcal{N}$ of coefficients $C_{ij}^k$ for all metric functions, since this will impact on the portability of this approximation. If the function $F_k$ with $k\in\{0,1,2\}$ considers $N_k$ polynomials basis, then total number of coefficients $\mathcal{N}$ will be $\mathcal{N}= n\times (N_0+N_1+N_2)$.\\

In this study we have set ${n}=7$, $\{N_0,N_1,N_2\}=\{34,34,40\}$ for region $A$, which corresponds to 756 coefficients $C_{ij}^k$ in total, for all metric functions. Similarly, for region $B$ we have $n=4$ and $\{N_0,N_1,N_2\}=\{34,34,40\}$, with 432 coefficients $C_{ij}^k$ in total. These numbers are explicitly detailed in Appendix~\ref{sec:appendix2} and Appendix~\ref{sec:appendix3} {simply to show that, albeit a large set of numbers, they can still be displayed in a few pages}. However, for practical purposes, it is also possible to access the coefficients $C_{ij}^k$ for each region in the website link~\cite{weblinkSpacetimeData}, together with a \textsc{mathematica} Notebook to quickly read and process the data.

The reported level of numerical accuracy comes at the cost of dealing with a large number of coefficients. However, if one is willing to accept potentially larger violations of the field equations, then it is possible to reduce drastically the number of coefficients. Indeed, the maximum error $\epsilon_{\textrm{max}}$ for all field equations within the entire region $A$ decays approximately\footnote{This approximation starts to break down for $\mathcal{N}$ larger than $\sim$ 700 (smaller than $\sim$ 30) due to other scaling effects.} with the exponential of the total number of coefficients $\mathcal{N}$:
\[\mathcal{O}\left(\epsilon_{\textrm{max}}\right) \sim 10^{-1.8 -\mathcal{N}/203}\,.\]
Hence having only $\mathcal{N}=$40 coefficients would lead to an accuracy of order $\sim 10^{-2}$. A similar scaling law also applies to region $B$.

Although this approximation involves hundreds of parameters, it can still be considered relatively compact. To illustrate this, let us examine the storage space required for a typical text file containing all 1188 coefficients ($432 + 756$) for regions $A$ and $B$, which amounts to approximately 30 kB. Now, let us contrast this with an alternative scenario: saving 300 individual solutions in the same region of the solution space. Each solution includes 500 data points along the radial coordinate for each metric function. This alternative setup would occupy around 9 MB, which is nearly 300 times larger than the memory size required for the first scenario. Furthermore, in addition to occupying a larger memory space, storing solution data at discrete points would still require employing interpolation methods to bridge the gap to the continuum. In contrast, the approach outlined herein offers an analytical expansion upfront, eliminating the necessity for further interpolations. 

From these results, it is clear that by integrating Acceleration-Informed and Standard methods, we can achieve a reliable analytical approximation for some sections of the BS family. However, it is important to note that expanding the domain of these regions tends to rapidly reduce the accuracy of these approximations. Specifically, BSs close to the Newtonian limit, where $\omega/\mu\to 1$, tend to be less compact and prove to be more challenging to efficiently approximate within the methodology applied in this study. This often resulted in much larger violations of the field equations when compared to regions $A$ and $B$. A similar problem also arises in regions to the left of $A$, although these are more compact. Both issues might be connected to the existence of larger metric gradients, defined with respect to the compactified radial coordinate $x$. It was not possible to meet this challenge by simply increasing the total number of coefficients, which suggests that further improving the results on this paper requires using enhanced numerical methods with higher numerical precision.

\section{Conclusions}\label{sec:conclusions}

 Approximations of specific spacetime solutions are common in the literature - see $e.g.$~\cite{Cardoso:2014rha,Johannsen:2011dh,Rezzolla:2014mua,Mukazhanov:2024rka,Pappas:2023mbp,Konoplya:2020hyk,Konoplya:2016jvv}. However, similar approximations applied to an entire family of solutions seem to be lacking. This paper provides an exploratory study in this direction, showcasing how an approximation of an entire family of BS solutions can be achieved in terms of a double expansion of three field functions on a Chebyshev-type basis. Such an approximation was found to be competitive in terms of usability of the data and in reducing storage size. \\

The current paper also aims to make BS star numerical data easily accessible. To this end, appropriate notebooks with the coefficients of the double expansion were made publicly available. This will allow the community to use the BS (sub-)family discussed here, without any further processing required. In a more sophisticated presentation, one could envisage an online  platform that not only provides the necessary coefficients for a chosen solution but simultaneously includes visualization and analysis tools. Work in this direction is underway.\\

Despite the promising results in this paper, there is still room to further refine this methodology in future work:

 \begin{itemize}
    \item The approximation in this paper is restricted to two regions $A$ and $B$ that do not cover several members of the BS family. In addition, the current methodology does not converge as quickly outside of these regions, since BS solutions become either very compact (to the left of $A$) or very diluted (to the right of region $B$), which is translated in the need to approximate larger gradients. This issue might be addressed by introducing a new definition of the radial coordinate with a different method of compactifying infinity to a finite coordinate. This approach could better accommodate solutions with larger frequency (i.e. in the Newtonian limit), with $\omega/\mu\simeq 1$, which are characterized by more dilute distributions of the scalar field that cover greater radial distances. We note however that as one approaches the Newtonian limit there is a scaling symmetry that allows covering that branch with a single solution~\cite{Annulli:2020lyc}.
    \item The results in this paper can be further improved by implementing more sophisticated numerical techniques to enhance precision. Some examples are advanced methods for numerical integration with Gaussian quadrature tailored for the specific integrals encountered. In this analysis, the majority of calculations were conducted utilizing built-in functions within Wolfram \textsc{mathematica}. While these functions are robust and versatile for various applications, they may not achieve the maximum numerical precision permitted by the allocated number of digits.
\end{itemize}

\bigskip

\section*{Acknowledgements}
The authors would like to thank Eugen Radu for his valuable comments and discussions.\\
This work is supported by the Center for Research and Development in Mathematics and Applications (CIDMA) through the Portuguese Foundation for Science and Technology (Funda\c c\~ao para a Ci\^encia e a Tecnologia), UIDB/04106/2020, UIDP/04106/2020,
https://doi.org/10.54499/UIDB/04106/2020 \\
and https://doi.org/10.54499/UIDP/04106/2020.
The authors also acknowledge support from the two projects
http://doi.org/10.54499/PTDC/FISAST/3041/2020,
http://doi.org/10.54499/CERN/FIS-PAR/0024/2021 and https://doi.org/10.54499/2022.04560.PTDC.  This work has further been supported by the European Horizon Europe staff exchange (SE) programme HORIZON-MSCA2021-SE-01 Grant No. NewFunFiCO-101086251. PC is supported by the Individual CEEC program\\ http://doi.org/10.54499/2020.01411.CEECIND/CP1589/CT0035 of 2020, funded by the Portuguese Foundation for Science and Technology (Funda\c c\~ao para a Ci\^encia e a Tecnologia).
Jianzhi Yang is supported by the China Scholarship Council.

\appendix
\section{Coefficients for Boson Star solution with maximum mass}\label{sec:appendix}

\begin{table}[H]
\centering
\begin{tabular}{|c|c|c|c|}
\hline
&  $\alpha_i^{(0)}$ ($F_0$) & $\alpha_i^{(1)}$ ($F_1$) & $\alpha_i^{(2)}$ ($F_2 \equiv \phi$)\\
 \hline
$i=0$ & -0.253301 &  0.21521 & 0.117883 \\
$i=1$ & 0.188197 &  -0.149873 & -0.108147 \\
$i=2$ & 0.0739083 & -0.0660349 & -0.033298 \\
$i=3$ & 0.00470212 & -0.0102588 & 0.00895514\\
$i=4$ &-0.0101331 & 0.00632009 & 0.0134482\\
$i=5$ & -0.00495984 & 0.00494095 & 0.00529592\\
$i=6$ & -6.72861$\times 10^{-7}$ & 0.00106567 & -0.000679038\\
\hline
\end{tabular}
\caption{Coefficients of the BS solution with maximal mass, obtained using the ``Standard Method'' described in Section~\ref{secApprox}.}
\label{table:data1}
\end{table}

\begin{table}[H]
\centering
\begin{tabular}{|c|c|c|c|}
\hline
&  $\alpha_i^{(0)}$ ($F_0$) & $\alpha_i^{(1)}$ ($F_1$) & $\alpha_i^{(2)}$ ($F_2 \equiv \phi$)\\
 \hline
  $i=0$ & -0.253301 & 0.21521 & 0.117883\\
 $i=1$ & 0.188197 & -0.149873 & -0.108147\\
 $i=2$ & 0.0739083 & -0.0660349  & -0.033298\\
 $i=3$ & 0.00470212 & -0.0102588 &  0.00895514\\
 $i=4$ &-0.0101331 & 0.00632009 & 0.0134482\\
 $i=5$ & -0.00495984 & 0.00494095 & 0.00529592\\
 $i=6$ & -6.72864$\times 10^{-7}$ & 0.00106567 & -0.000679038\\
\hline
\end{tabular}
\caption{Coefficients of the BS solution with maximal mass, obtained using the ``Acceleration-Informed Method'' described in Section~\ref{secApprox}.}
\label{table:data2}
\end{table}

\section{Coefficients for region $A$}\label{sec:appendix2}
\begin{table}[H]
\centering
\begin{tabular}{|c|c|c|c|c|}
\hline
& & region $A$ & region $A$ & region $A$\\\hline
 & & $C_{ij}^{0}$ ($F_0$) & $C_{ij}^{1}$ ($F_1$) & $C_{ij}^{2}$ ($F_2 \equiv \phi$)\\\hline$i=0$ & $j=0$ & $-3.116009898971708 \times 10^{-1}$ & $2.528251713152068 \times 10^{-1}$ & $1.4275684107592643 \times 10^{-1}$ \\
$i=0$ & $j=1$ & $5.479818831274786 \times 10^{-2}$ & $-3.4200628203106835 \times 10^{-2}$ & $-2.3171285232899934 \times 10^{-2}$ \\
$i=0$ & $j=2$ & $-1.8094595336779917 \times 10^{-3}$ & $2.0000169599540167 \times 10^{-4}$ & $5.969580626271306 \times 10^{-4}$ \\
$i=0$ & $j=3$ & $2.0616572146105305 \times 10^{-4}$ & $-6.495762618364054 \times 10^{-5}$ & $-7.513857634559049 \times 10^{-5}$ \\
$i=0$ & $j=4$ & $-2.9432079602786443 \times 10^{-5}$ & $9.94494103030635 \times 10^{-6}$ & $1.0847427537596643 \times 10^{-5}$ \\
$i=0$ & $j=5$ & $4.735654018588676 \times 10^{-6}$ & $-1.6537426538053208 \times 10^{-6}$ & $-1.7583587788623565 \times 10^{-6}$ \\
\hline
\end{tabular}
\caption{Coefficients $C_{ij}^k$ for region $A$, discussed in Section~\ref{sec:ApproxFamily}.}
\label{table:Data0}
\end{table}
\begin{table}[H]
\centering
\begin{tabular}{|c|c|c|c|c|}
\hline
& & region $A$ & region $A$ & region $A$\\\hline
 & & $C_{ij}^{0}$ ($F_0$) & $C_{ij}^{1}$ ($F_1$) & $C_{ij}^{2}$ ($F_2 \equiv \phi$)\\\hline$i=5$ & $j=5$ & $1.4429402345898943 \times 10^{-6}$ & $-2.77990818046908 \times 10^{-7}$ & $-6.619693959420983 \times 10^{-7}$ \\
$i=5$ & $j=6$ & $-2.2705506647132156 \times 10^{-7}$ & $4.482573137911994 \times 10^{-8}$ & $1.024985574668279 \times 10^{-7}$ \\
$i=6$ & $j=0$ & $1.5986536177770667 \times 10^{-3}$ & $-3.95055568845098 \times 10^{-4}$ & $-2.4824129152094905 \times 10^{-3}$ \\
$i=6$ & $j=1$ & $-1.0862972142134741 \times 10^{-3}$ & $1.2447681965661184 \times 10^{-3}$ & $1.3282283092701028 \times 10^{-3}$ \\
$i=6$ & $j=2$ & $-3.7865625174634427 \times 10^{-4}$ & $1.0731172966901583 \times 10^{-4}$ & $2.9968852022764363 \times 10^{-4}$ \\
$i=6$ & $j=3$ & $7.664168876743927 \times 10^{-5}$ & $-3.6494263140370244 \times 10^{-5}$ & $-4.900698855812443 \times 10^{-5}$ \\
\hline
\end{tabular}
\caption{Coefficients $C_{ij}^k$ for region $A$, discussed in Section~\ref{sec:ApproxFamily}.}
\label{table:Data1}
\end{table}
\begin{table}[H]
\centering
\begin{tabular}{|c|c|c|c|c|}
\hline
& & region $A$ & region $A$ & region $A$\\\hline
 & & $C_{ij}^{0}$ ($F_0$) & $C_{ij}^{1}$ ($F_1$) & $C_{ij}^{2}$ ($F_2 \equiv \phi$)\\\hline$i=11$ & $j=3$ & $3.902217029400723 \times 10^{-6}$ & $-3.499950012419509 \times 10^{-6}$ & $-4.713278289886238 \times 10^{-6}$ \\
$i=11$ & $j=4$ & $1.11923648892194 \times 10^{-6}$ & $-1.2257812102232468 \times 10^{-7}$ & $-4.692304769435028 \times 10^{-7}$ \\
$i=11$ & $j=5$ & $-1.9809767931510514 \times 10^{-7}$ & $7.695554640957135 \times 10^{-8}$ & $1.2397324027407781 \times 10^{-7}$ \\
$i=11$ & $j=6$ & $2.267603989912001 \times 10^{-9}$ & $-3.1482108400725086 \times 10^{-9}$ & $-1.2746649403958987 \times 10^{-9}$ \\
$i=12$ & $j=0$ & $-2.5530808482339578 \times 10^{-6}$ & $2.165607664047255 \times 10^{-5}$ & $1.244994731703918 \times 10^{-4}$ \\
$i=12$ & $j=1$ & $-8.688814358498318 \times 10^{-6}$ & $2.8884600909426568 \times 10^{-5}$ & $6.242722887646972 \times 10^{-5}$ \\
\hline
\end{tabular}
\caption{Coefficients $C_{ij}^k$ for region $A$, discussed in Section~\ref{sec:ApproxFamily}.}
\label{table:Data2}
\end{table}
\begin{table}[H]
\centering
\begin{tabular}{|c|c|c|c|c|}
\hline
& & region $A$ & region $A$ & region $A$\\\hline
 & & $C_{ij}^{0}$ ($F_0$) & $C_{ij}^{1}$ ($F_1$) & $C_{ij}^{2}$ ($F_2 \equiv \phi$)\\\hline$i=17$ & $j=1$ & $4.980813109050664 \times 10^{-7}$ & $-1.6714510901976345 \times 10^{-6}$ & $-7.222214837941302 \times 10^{-6}$ \\
$i=17$ & $j=2$ & $2.1851526955373015 \times 10^{-7}$ & $-5.234342717025583 \times 10^{-7}$ & $2.493467477191691 \times 10^{-7}$ \\
$i=17$ & $j=3$ & $1.0444309145582003 \times 10^{-8}$ & $-5.502827927625359 \times 10^{-8}$ & $2.0531443169570305 \times 10^{-7}$ \\
$i=17$ & $j=4$ & $2.4873509362074873 \times 10^{-8}$ & $-1.2410719855570904 \times 10^{-10}$ & $1.080552518925082 \times 10^{-8}$ \\
$i=17$ & $j=5$ & $7.10014547871624 \times 10^{-9}$ & $2.6222393451444064 \times 10^{-9}$ & $-1.1320879346430177 \times 10^{-9}$ \\
$i=17$ & $j=6$ & $-2.136457074760395 \times 10^{-9}$ & $7.927320093873406 \times 10^{-10}$ & $1.0668799526355317 \times 10^{-9}$ \\
\hline
\end{tabular}
\caption{Coefficients $C_{ij}^k$ for region $A$, discussed in Section~\ref{sec:ApproxFamily}.}
\label{table:Data3}
\end{table}
\begin{table}[H]
\centering
\begin{tabular}{|c|c|c|c|c|}
\hline
& & region $A$ & region $A$ & region $A$\\\hline
 & & $C_{ij}^{0}$ ($F_0$) & $C_{ij}^{1}$ ($F_1$) & $C_{ij}^{2}$ ($F_2 \equiv \phi$)\\\hline$i=22$ & $j=6$ & $-4.5579155894828185 \times 10^{-11}$ & $-9.673618416632944 \times 10^{-12}$ & $7.78527612736465 \times 10^{-11}$ \\
$i=23$ & $j=0$ & $4.282375580292912 \times 10^{-8}$ & $-5.440816156497941 \times 10^{-8}$ & $1.83727811005679 \times 10^{-6}$ \\
$i=23$ & $j=1$ & $1.3111648190927538 \times 10^{-8}$ & $2.9378491339129815 \times 10^{-8}$ & $9.75595184959379 \times 10^{-7}$ \\
$i=23$ & $j=2$ & $-1.4067399106360416 \times 10^{-8}$ & $3.5958636634991463 \times 10^{-8}$ & $-1.523188686825082 \times 10^{-8}$ \\
$i=23$ & $j=3$ & $-6.21675605202314 \times 10^{-9}$ & $8.745165901251907 \times 10^{-9}$ & $-3.1644705217977216 \times 10^{-8}$ \\
$i=23$ & $j=4$ & $-1.0205859342249992 \times 10^{-9}$ & $6.764040900799283 \times 10^{-10}$ & $-4.7152402059381355 \times 10^{-9}$ \\
\hline
\end{tabular}
\caption{Coefficients $C_{ij}^k$ for region $A$, discussed in Section~\ref{sec:ApproxFamily}.}
\label{table:Data4}
\end{table}
\begin{table}[H]
\centering
\begin{tabular}{|c|c|c|c|c|}
\hline
& & region $A$ & region $A$ & region $A$\\\hline
 & & $C_{ij}^{0}$ ($F_0$) & $C_{ij}^{1}$ ($F_1$) & $C_{ij}^{2}$ ($F_2 \equiv \phi$)\\\hline$i=28$ & $j=4$ & $1.4448044981295953 \times 10^{-10}$ & $-1.1470912591107605 \times 10^{-10}$ & $9.23937310586122 \times 10^{-10}$ \\
$i=28$ & $j=5$ & $6.059753969188389 \times 10^{-12}$ & $1.305137861541355 \times 10^{-11}$ & $-4.732787622190954 \times 10^{-12}$ \\
$i=28$ & $j=6$ & $-9.814648473328182 \times 10^{-13}$ & $4.6030358147692105 \times 10^{-12}$ & $-1.0410386062937564 \times 10^{-11}$ \\
$i=29$ & $j=0$ & $-5.0147144507682855 \times 10^{-9}$ & $8.547597826457306 \times 10^{-9}$ & $-2.627276683804707 \times 10^{-7}$ \\
$i=29$ & $j=1$ & $-4.094455714725037 \times 10^{-9}$ & $3.4130073519437567 \times 10^{-9}$ & $-1.1772042599323504 \times 10^{-7}$ \\
$i=29$ & $j=2$ & $2.5994956648142344 \times 10^{-10}$ & $-2.133939738714026 \times 10^{-9}$ & $1.318887138872926 \times 10^{-8}$ \\
\hline
\end{tabular}
\caption{Coefficients $C_{ij}^k$ for region $A$, discussed in Section~\ref{sec:ApproxFamily}.}
\label{table:Data5}
\end{table}
\begin{table}[H]
\centering
\begin{tabular}{|c|c|c|c|c|}
\hline
& & region $A$ & region $A$ & region $A$\\\hline
 & & $C_{ij}^{0}$ ($F_0$) & $C_{ij}^{1}$ ($F_1$) & $C_{ij}^{2}$ ($F_2 \equiv \phi$)\\\hline$i=34$ & $j=2$ &   &   & $-1.1657062340421034 \times 10^{-8}$ \\
$i=34$ & $j=3$ &   &   & $-2.054838809443118 \times 10^{-9}$ \\
$i=34$ & $j=4$ &   &   & $-1.2640417979652108 \times 10^{-10}$ \\
$i=34$ & $j=5$ &   &   & $1.2421473667049598 \times 10^{-11}$ \\
$i=34$ & $j=6$ &   &   & $2.7981772495195987 \times 10^{-12}$ \\
$i=35$ & $j=0$ &   &   & $3.5935928677783697 \times 10^{-8}$ \\
\hline
\end{tabular}
\caption{Coefficients $C_{ij}^k$ for region $A$, discussed in Section~\ref{sec:ApproxFamily}.}
\label{table:Data6}
\end{table}

\section{Coefficients for region $B$}\label{sec:appendix3}
\begin{table}[H]
\centering
\begin{tabular}{|c|c|c|c|c|}
\hline
& & region $B$ & region $B$ & region $B$\\\hline
 & & $C_{ij}^{0}$ ($F_0$) & $C_{ij}^{1}$ ($F_1$) & $C_{ij}^{2}$ ($F_2 \equiv \phi$)\\\hline$i=0$ & $j=0$ & $-2.1177964205296655 \times 10^{-1}$ & $1.8459356588027727 \times 10^{-1}$ & $9.948374570613087 \times 10^{-2}$ \\
$i=0$ & $j=1$ & $4.593800065129586 \times 10^{-2}$ & $-3.430326939968538 \times 10^{-2}$ & $-2.043332815152876 \times 10^{-2}$ \\
$i=0$ & $j=2$ & $-6.693943329662073 \times 10^{-4}$ & $-1.4118531297404455 \times 10^{-4}$ & $1.8346537782522443 \times 10^{-4}$ \\
$i=0$ & $j=3$ & $4.142110064830607 \times 10^{-5}$ & $-1.1761235493549527 \times 10^{-5}$ & $-1.513203152777838 \times 10^{-5}$ \\
$i=1$ & $j=0$ & $1.4615361357956225 \times 10^{-1}$ & $-1.2026699725539867 \times 10^{-1}$ & $-8.563833207759314 \times 10^{-2}$ \\
$i=1$ & $j=1$ & $-4.554783083242184 \times 10^{-2}$ & $3.2481158344359073 \times 10^{-2}$ & $2.456131829488144 \times 10^{-2}$ \\
\hline
\end{tabular}
\caption{Coefficients $C_{ij}^k$ for region $B$, discussed in Section~\ref{sec:ApproxFamily}.}
\label{table:Data7}
\end{table}
\begin{table}[H]
\centering
\begin{tabular}{|c|c|c|c|c|}
\hline
& & region $B$ & region $B$ & region $B$\\\hline
 & & $C_{ij}^{0}$ ($F_0$) & $C_{ij}^{1}$ ($F_1$) & $C_{ij}^{2}$ ($F_2 \equiv \phi$)\\\hline$i=10$ & $j=0$ & $4.756664770992647 \times 10^{-5}$ & $-1.3333684063495321 \times 10^{-4}$ & $-2.3330175281656618 \times 10^{-4}$ \\
$i=10$ & $j=1$ & $2.0006101668032903 \times 10^{-4}$ & $-1.7118519193737731 \times 10^{-4}$ & $-3.6660814986533734 \times 10^{-4}$ \\
$i=10$ & $j=2$ & $1.716181201689821 \times 10^{-5}$ & $4.728581415445831 \times 10^{-6}$ & $-8.73575248184694 \times 10^{-6}$ \\
$i=10$ & $j=3$ & $-9.65209787108502 \times 10^{-6}$ & $6.95012951781177 \times 10^{-6}$ & $1.0896638961866076 \times 10^{-5}$ \\
$i=11$ & $j=0$ & $-4.925557288070077 \times 10^{-5}$ & $1.0633207984198905 \times 10^{-5}$ & $8.030673454671478 \times 10^{-5}$ \\
$i=11$ & $j=1$ & $6.685851143621083 \times 10^{-5}$ & $-9.983318769769552 \times 10^{-5}$ & $-2.0340531664468147 \times 10^{-4}$ \\
\hline
\end{tabular}
\caption{Coefficients $C_{ij}^k$ for region $B$, discussed in Section~\ref{sec:ApproxFamily}.}
\label{table:Data8}
\end{table}
\begin{table}[H]
\centering
\begin{tabular}{|c|c|c|c|c|}
\hline
& & region $B$ & region $B$ & region $B$\\\hline
 & & $C_{ij}^{0}$ ($F_0$) & $C_{ij}^{1}$ ($F_1$) & $C_{ij}^{2}$ ($F_2 \equiv \phi$)\\\hline$i=20$ & $j=0$ & $-2.0597961741180693 \times 10^{-7}$ & $-1.5871649342182277 \times 10^{-7}$ & $-9.914418642421982 \times 10^{-6}$ \\
$i=20$ & $j=1$ & $7.371899596227511 \times 10^{-7}$ & $-1.6981289410065852 \times 10^{-6}$ & $-7.835515101597688 \times 10^{-6}$ \\
$i=20$ & $j=2$ & $5.286137666660203 \times 10^{-7}$ & $-6.835425874872221 \times 10^{-7}$ & $-2.12123886435416 \times 10^{-7}$ \\
$i=20$ & $j=3$ & $1.2747981301109312 \times 10^{-7}$ & $-5.743478837552197 \times 10^{-8}$ & $2.3594107628239738 \times 10^{-7}$ \\
$i=21$ & $j=0$ & $-5.254921503032333 \times 10^{-7}$ & $5.608155611275032 \times 10^{-7}$ & $-4.524732181136695 \times 10^{-6}$ \\
$i=21$ & $j=1$ & $-7.5899853301347555 \times 10^{-8}$ & $-4.244970468885181 \times 10^{-7}$ & $-6.578249331661877 \times 10^{-6}$ \\
\hline
\end{tabular}
\caption{Coefficients $C_{ij}^k$ for region $B$, discussed in Section~\ref{sec:ApproxFamily}.}
\label{table:Data9}
\end{table}
\begin{table}[H]
\centering
\begin{tabular}{|c|c|c|c|c|}
\hline
& & region $B$ & region $B$ & region $B$\\\hline
 & & $C_{ij}^{0}$ ($F_0$) & $C_{ij}^{1}$ ($F_1$) & $C_{ij}^{2}$ ($F_2 \equiv \phi$)\\\hline$i=30$ & $j=0$ & $1.2351076478785172 \times 10^{-8}$ & $-2.3495145474671095 \times 10^{-8}$ & $-1.767165276202997 \times 10^{-7}$ \\
$i=30$ & $j=1$ & $3.337618488464078 \times 10^{-8}$ & $-5.3076890094034255 \times 10^{-8}$ & $2.816209739925302 \times 10^{-7}$ \\
$i=30$ & $j=2$ & $1.3693294257664912 \times 10^{-8}$ & $-1.6362408561680504 \times 10^{-8}$ & $1.4429481712513186 \times 10^{-7}$ \\
$i=30$ & $j=3$ & $1.3561631069879905 \times 10^{-9}$ & $-1.7889855891198339 \times 10^{-10}$ & $1.97831795505415 \times 10^{-8}$ \\
$i=31$ & $j=0$ & $3.890594812314744 \times 10^{-10}$ & $-4.22710655097792 \times 10^{-9}$ & $-2.7366789604258086 \times 10^{-7}$ \\
$i=31$ & $j=1$ & $1.5379623078723237 \times 10^{-8}$ & $-2.915356619925415 \times 10^{-8}$ & $5.426272919613457 \times 10^{-8}$ \\
\hline
\end{tabular}
\caption{Coefficients $C_{ij}^k$ for region $B$, discussed in Section~\ref{sec:ApproxFamily}.}
\label{table:Data10}
\end{table}

\bibliographystyle{ieeetr}  
\bibliography{references}

\end{document}